\documentclass[fleqn,usenatbib]{mnras}

\usepackage{newtxtext,newtxmath}

\usepackage[T1]{fontenc}
\DeclareRobustCommand{\VAN}[3]{#2}
\let\VANthebibliography\thebibliography
\def\thebibliography{\DeclareRobustCommand{\VAN}[3]{##3}\VANthebibliography}

\usepackage{graphicx}	
\usepackage{amsmath}	
\usepackage{widetext}
\usepackage[usenames,dvipsnames]{xcolor}
\usepackage{orcidlink}
\usepackage{multirow} 

\title[Can $f(T)$ models play a bridge between early and late time Universe?]{Can teleparallel $f(T)$ models play a bridge between early and late time Universe?}

\author[N. S. Kavya et al.]{
N. S. Kavya$^{\orcidlink{0000-0001-8561-130X}1}$\thanks{E-mail: kavya.samak.10@gmail.com},
Sai Swagat Mishra$^{\orcidlink{0000-0003-0580-0798}2}$\thanks{E-mail: saiswagat009@gmail.com},
P.K. Sahoo$^{\orcidlink{0000-0003-2130-8832}2}$ \thanks{E-mail: pksahoo@hyderabad.bits-pilani.ac.in},
V. Venkatesha$^{\orcidlink{0000-0002-2799-2535}1}$\thanks{E-mail: vensmath@gmail.com}
\\
$^{1}$Department of P.G. Studies and Research in Mathematics,
 Kuvempu University, Shankaraghatta, Shivamogga 577451, Karnataka, India\\
$^{2}$Department of Mathematics, Birla Institute of Technology and
Science-Pilani, Hyderabad Campus, Hyderabad-500078, India.}

\date{Accepted XXX. Received YYY; in original form ZZZ}

\pubyear{2024}
\begin{document}
\label{firstpage}
\pagerange{\pageref{firstpage}--\pageref{lastpage}}
\maketitle

\begin{abstract}
The ability of Big Bang Nucleosynthesis theory to accurately predict the primordial abundances of helium and deuterium, as well as the baryon content of the Universe, is considered one of the most significant achievements in modern physics. In the present study, we consider two highly motivated hybrid $f(T)$ models and constrain them using the observations from the Big Bang Nucleosynthesis era. In addition, using late-time observations of Cosmic Chronometers and Gamma-Ray-Bursts, the ranges of the model parameters are confined which are in good agreement with early time bounds. Subsequently, the common ranges obtained from the analysis for early and late time are summarized. Further, we verify the intermediating epochs by investigating the profiles of cosmographic parameters using the model parameter values from the common range. From this study, we find the considered teleparallel models are viable candidates to explain the primordial-intermediating-present epochs.
\end{abstract}

\section{Introduction}

The present understanding of the Universe's evolution is strongly explained on the basis of the Friedmann-Robertson-Walker (FRW) cosmological model, commonly known as the hot Big Bang model. Cosmological observations from Type Ia Supernovae \citep{SupernovaSearchTeam:1998fmf,SupernovaCosmologyProject:1998vns,SupernovaSearchTeam:2004lze}, Baryon acoustic oscillations (BAO) \citep{SDSS:2005xqv,SDSS:2009ocz}, galaxy redshift survey \citep{Fedeli:2008fh}, large-scale structure (LSS) \citep{Koivisto:2005mm,Daniel:2008et,Nadathur:2020kvq}, and cosmic microwave background radiation (CMBR) \citep{WMAP:2003elm}, prove that the Universe is undergoing accelerated expansion. This phenomenon is generally attributed to the presence of a dark energy (DE) sector in the Universe, a hypothetical energy source characterized by negative pressure. The most successful model to explain this phenomenon is the one in which the cosmological constant  $\Lambda $ explains DE, known as the concordance model. The concrete affirmation and validity of the model comes from the very beginning of the timeline covering primordial nucleosynthesis to the present-time acceleration. Based on the cosmological principle and the high energy interactions of fundamental elements, the standard cosmological model is capable of predicting the events that can reach up to nearly $10^{-43}$ s of the post-Big Bang era.  Currently, the standard model of particle physics, represented by the $SU(3)_C \otimes SU(2)_L \otimes U(1)_Y$ gauge theory of strong and electroweak interactions, presents a fundamental framework for understanding quarks and leptons which has been experimentally verified at energies up to approximately 1000 GeV. 

Numerous observational surveys, with several teams working together, probe the evolution of the Universe at different stages, covering a wide range of observational factors. For instance, the Hubble parameter $H_0$, the present mass density $\rho_0$, the deceleration parameter $q_0$, and the distance modulus $\mu$ provide insight into late-time phenomena of the Universe \citep{WMAP:2006bqn,WMAP:2010qai,Demianski:2016zxi}, and observations of the transition redshift $z_t$ are crucial for understanding changes in the Universe's phase. However, observations of the intermediate epoch of the Universe remain under explored to date. Interestingly, there are several significant experimental observations of the primordial Universe to center on, such as the abundance of $^3$He, $^4$He, D, $^7$Li, the baryon-to-photon ratio, and the CMBR, which help us predict dynamics as early as 0.01 s after the Big Bang. Since the Universe undergoes expansion and cooling in its early stages, the rates of weak interactions, pivotal for maintaining thermal equilibrium between neutrinos and matter, decrease progressively. Likewise, the rates of reactions converting neutrons and protons decline. Eventually, these rates dip below the expansion rate, leading to a freeze-out. Simultaneously, the strong and electromagnetic nuclear reaction rate also decreases and undergoes freeze-out. These epochs contribute to the nucleosynthesis era of the Big Bang model \citep{Fields:2019pfx,Barrow:2020kug,Anagnostopoulos:2022gej,Kang:2008zi,Mishra:2023onl,Capozziello:2017bxm}. 

Though the standard cosmological model provides viable explanations for phenomena from early to late times and aligns well with observational data, it has some limitations \citep{Joyce:2014kja,Tsujikawa:2003jp,Velten:2014nra,Astesiano:2022ozl,Katsuragawa:2016yir,Zaregonbadi:2016xna,Joudaki:2016kym}. This necessitates exploring models beyond $\Lambda$CDM. While quintessence models, which incorporate scalar fields with potential, can explain cosmological data with certain adjustments, the absence of strongly motivated scalar field models from theoretical particle physics is significant. Consequently, the effects of cosmic acceleration might be better explained within the framework of modified gravity theories, where extensions of General Relativity (GR) are considered.

In this study, we explore an extension of gravitational theory known as the Teleparallel Equivalent of General Relativity (TEGR), which was initially proposed by Einstein himself in \citep{Unzicker:2005in, Maluf:2002zc, Obukhov:2002tm, Ferraro:2006jd, Linder:2010py}. TEGR describes the gravitational field using the torsion tensor. Although many results from TEGR match those from GR, extensions of TEGR can yield different outcomes compared to extensions of GR \citep{Cai:2015emx}. As a result, torsion-based modified theories have garnered significant attention and have been thoroughly investigated in recent years \citep{Wu:2010mn,Chen:2010va,Dent:2010nbw,Sudharani:2023hss,Bahamonde:2021srr,Moreira:2021xfe,Mishra:2023khd,Kofinas:2014daa,Gonzalez-Espinoza:2021mwr}. One can also see some interesting works that adopt different techniques to constrain $f(T)$ gravity models \citep{Nunes:2016qyp,Nunes:2016plz,Ren:2022aeo,Briffa:2020qli}. Motivated by this, we study two teleparallel $f(T)$ models and their influence on the dynamics of the Universe from early to late times. 

Article organization: In \autoref{sec2} we provide a detailed geometric foundation of $f(T)$ teleparallel gravity. The \autoref{sec3} presents the formalism of BBN. Using BBN constraints we analyze two hybrid gravity models in \autoref{sec4}. The early Universe constraints are obtained from BBN, while the late time constraints are obtained from recent observational data sets of Cosmic chronometers and Gamma-Ray-Bursts. The methodology and results of this procedure are explained in \autoref{sec5} and in \autoref{sec6}, we conclude our results. 

\section{Formalism of $f(T)$ gravity}\label{sec2}
The simplest departure from the Teleparallel Equivalent of General Relativity is considering an arbitrary functional form of the torsion scalar gives rise to the so-called "$f(T)$ theory". This theory uses the Weitzenb\"{o}ck connection instead of the usual torsion-less Levi-Civita connection of GR. Using the connection, the non-null torsion tensor can be obtained as
\begin{equation}\label{eq:torsiontensor}
    T_{\mu \nu}^{\lambda}=\overset{w}{\Gamma}_{\nu \mu}^\lambda-\overset{w}{\Gamma}_{\mu \nu}^\lambda=e_{\gamma}^\lambda(\partial_{\mu}e_{\nu}^\gamma-\partial_{\nu}e_{\mu}^\gamma).
\end{equation}
Here $\overset{w}{\Gamma}_{\nu \mu}^\lambda$ is the Weitzenb\"{o}ck connection and the tetrad fields $e_{\mu}^\gamma$ or $e_{\mu}(x^\gamma)$ create a tangent space at each point $x^\gamma$ of the manifold. The metric tensor of the manifold can be written in terms of the tetrad fields as $g_{\mu \nu}(x)=\eta_{\alpha \beta} \, e_\mu^{\alpha}(x) \, e_\nu^{\beta}(x)$, with the pseudo-Riemannian metric $\eta_{\alpha \beta}=diag(1,-1,-1,-1)$.

 The torsion scalar can be defined as 
\begin{equation}\label{eq:torsionscalar}
    T\equiv {S_{\lambda}}^{\mu \nu} T_{\mu \nu}^\lambda
\end{equation}
where the superpotential tensor is defined as ${S_{\lambda}}^{\mu \nu} \equiv \frac{1}{2}({K^{\mu \nu}}_{\lambda}+\delta_\lambda^\mu {T^{\alpha \nu}}_{\alpha}-\delta_\lambda^\nu {T^{\alpha \mu}}_{\alpha})$ and the contorsion tensor is defined as ${K^{\mu \nu}}_{\lambda} \equiv -\frac{1}{2}({T^{\mu \nu}}_{\lambda}-{T^{\nu \mu}}_{\lambda}-{T_{\lambda}}^{\mu \nu})$. Using the torsion scalar one can modify the GR action to teleparallel action as
\begin{equation}\label{eq:action}
    S=\frac{1}{2k^2} \int d^4xe[T+f(T)]
\end{equation}
where $k=\sqrt{8\pi G}$ and $e=det(e_\mu^A)=\sqrt{-g}$. 

Variation of the above action with respect to the tetrads gives rise to the motion equation
\begin{multline}
 \label{eq:field}
   (1+f_T)\left[e^{-1}\partial_\mu\,(e\,{e^\lambda_{\,A}} \,S_\lambda^{\,\, \nu \mu})-e^\alpha_{\,A} \,T^{\lambda}_{\,\, \mu \alpha}\, S_{\lambda}^{\,\, \mu \nu} \right]+\\
   e^{\lambda}_{\,A}\, S_\lambda^{\,\, \nu \mu}\, \partial_\mu T \, f_{TT}+\frac{1}{4} \,e^\nu_{\,A} \,[T+f]=4 \pi G \,e^\lambda_{\,A} \stackrel{em}{T}_\lambda^{\,\,\, \nu},
\end{multline}
 where $f_T, \, f_{TT}$ are the first and second derivative of $f$ with respect to the torsion scalar, respectively and $\stackrel{em}{T}_\lambda^{\,\,\, \nu}$ denotes the usual energy momentum tensor. For further discussion, we consider a spatially flat Friedmann-Lemaitre-Robertson-Walker (FLRW) metric of the form $ds^2=dt^2-a^2(t)\,\delta_{ij}\,dx^i\, dx^j$ with $a(t)$ being the scale factor in terms of time. The vierbein ansatz, corresponding to the metric, considered of the form $e_\mu^A=diag(1,a(t),a(t),a(t))$. The vierbein along with \eqref{eq:torsionscalar} gives rise to the relation $T=-6H^2$ where the Hubble parameter $H=\dot{a}/a$.

Inserting the vierbein in the motion equation \eqref{eq:field}, the modified Friedmann equations can be obtained as
\begin{equation}
\label{eq:motion1}
    12 H ^2 (1+f_T)+T+f=2k^2 \rho,
\end{equation}
\begin{equation}\label{eq:motion2}
    48 H ^2 \dot{H} f_{TT}-(1+f_T)(4\dot{H}+12H^2)+f-T=2 k^2 p,
\end{equation}
where $\rho=\rho_m+\rho_r$ and $p=p_m+p_r$. The above motion equations can be rewritten in the GR equivalent form as
\begin{gather}\label{eq:motiongr1}
    H^2=\frac{k^2}{3}(\rho +\rho_{DE}),\\ \label{eq:motiongr2}
    3H^2+2\dot{H}=-\frac{k^2}{3}(p +p_{DE}).
\end{gather}
 The dark energy components in the above equations are expressed in the context of $f(T)$ gravity as 
\begin{gather}\label{eq:rhode}
    \rho_{DE}=\frac{3}{k^2} \left(\frac{T f_T}{3}-\frac{f}{6}\right),\\\label{eq:pde}
    p_{DE}=\frac{1}{2k^2} \frac{f-T f_T +2 T^2 f_{TT}}{1+f_T+2 T f_{TT}}.
\end{gather}

\section{Background of BBN}\label{sec3}

Recent studies have suggested that BBN occurred within the initial fractions of seconds following the Big Bang, approximately around 0.01 s, extending to several hundred seconds after. During this period, the Universe was extremely hot and dense. BBN, along with the CMBR, provides firm validation for the high temperatures that characterized the primordial Universe. Here, we shall explore BBN physics in the context of $f(T)$ teleparallel gravity. The occurrence of BBN takes place in the era in which radiation was dominating. The relativistic elements (as well as mass-less radiation) in this stage have the energy density $\rho_r$ given by 

\begin{equation}\label{eq:rhor}
    \rho_r=\frac{\pi^2}{30} g_* \mathcal{T}^4.
\end{equation}

Here, $\mathcal{T}$ is the temperature, and $g_*$ represents the effective number of degrees of freedom, which is nearly equal to 10. The abundance of neutrons is calibrated by considering the proton-to-neutron conversion rate \citep{Kolb:1990vq}

\begin{equation}\label{eq:lambdapn}
    \lambda_{pn}(\mathcal{T}) = \lambda_{(n + \nu_e \rightarrow p + e^-)} + \lambda_{(n + e^+ \rightarrow p + \bar{\nu}_e)} + \lambda_{(n \rightarrow p + e^- + \bar{\nu}_e)}.
\end{equation}

The total rate is given by
\begin{equation}\label{lambdatot}
    \lambda_{tot}(\mathcal{T}) = \lambda_{pn}(\mathcal{T}) + \lambda_{np}(\mathcal{T}),
\end{equation}

where $\lambda_{np}(\mathcal{T})$ is the inverse of $\lambda_{pn}(\mathcal{T})$. In view of \eqref{eq:lambdapn}, the above equation takes the form \citep{Anagnostopoulos:2022gej}

\begin{equation}
   \lambda_{tot}(\mathcal{T})= 4 \mathcal{A} T 3(4!T 2 + 2 \times 3!QT + 2!Q2).
\end{equation}

In the above equation,  $Q$  represent the neutron-proton mass difference, given by  $Q = m_n - m_p = 1.29 \times 10^{-3} \, \text{GeV}$ . Additionally, $\mathcal{A}$ is taken as  $\mathcal{A} = 1.02 \times 10^{-11} \, \text{GeV}^{-4}$. Using the relation 
\begin{equation}
    Y_p \equiv e^{-\frac{t_n - \mathcal{T}_F}{\tau}}
 \frac{2e^{-\frac{Q}{T(\mathcal{T}_F)}}
}{1 + e^{-\frac{Q}{T(\mathcal{T}_F)}}
},
\end{equation}
the primordial mass fraction of $^4$He can be calibrated with $t_n$ and $\mathcal{T}_F$ being the freeze-out time and freeze-out temperature for nucleosynthesis and weak interactions, respectively, and $\tau$ refers to the mean lifetime of neutrons \citep{Barrow:2020kug}. 

In the context of GR, the first Friedmann equation can be written as $H^2=\frac{k^2}{3}\rho_{eff}$ \eqref{eq:motiongr1}, where $\rho_{eff}$ is the effective energy density of the system. Since radiation dominates in the BBN era, observational evidence indicates that any other contribution is negligible compared to radiation when the concordance model is considered. Thus, we have
\begin{equation}\label{eq:HGR}
    H^2 \approx \frac{k^2}{3} \rho_r \equiv H^2_{GR},
\end{equation}
where $k=1/M_p$, with $M_p = 1.22 \times 10^{19} \, \text{GeV}$ being the plank mass.

Here, we provide the convention adopted in the later part of this work: to distinguish the Hubble rate in GR from that obtained in $f(T)$ theory, we denote the former as $H_{GR}$ and the latter simply as $H$. Thus, equation \eqref{eq:motiongr1} along with the equation \eqref{eq:HGR} reads

\begin{equation}
    H = H_{GR} \sqrt{1 + \frac{\rho_{DE}}{\rho_r}}.
\end{equation}

Expanding the quantity within the square root to first order is justified by the dominance of dark energy density being significantly smaller than that of radiation during the radiation-dominated era and thus obtaining 

\begin{equation}\label{eq:deltaH}
    \Delta H = H - H_{GR} \approx \frac{\rho_{DE}}{\rho_r} \frac{H_{GR}}{2}.
\end{equation}

The equation \eqref{eq:rhor} in the above equation yields 

\begin{equation}\label{eq:H(T)}
    H(\mathcal{T}) = \sqrt{\frac{\pi^2 g_*}{90}} \frac{\mathcal{T}^2}{M_{p}}.
\end{equation}

Thus, the scale factor evolves as a $\sim t^{1/2}$, with $t$ being the cosmic time. The relation between temperature and time is therefore given by

\begin{equation}
    \frac{1}{t} \sim \left( \frac{32\pi^3 g_*}{90} \right)^{1/2} \frac{\mathcal{T}^2}{M_p}
\end{equation}

or in other words $\mathcal{T}(t) \sim (t/\text{sec})^{-1/2} \, \text{MeV}$. 

Central to BBN is the concept of the freeze-out temperature, $\mathcal{T}_F$. Neutrons and protons interconvert via weak interactions. When the temperature exceeds 1 MeV, much higher than the expansion rate, these conversions remain in equilibrium. However, as the temperature falls below 1 MeV, the neutron-to-proton ratio 'freezes out' at approximately 1/6, gradually declining due to the decay of free neutrons. The freeze-out temperature in hand is related to the specific Hubble rate  $H(\mathcal{T}_F) = \lambda_{tot}(\mathcal{T}_F) \approx c_q \mathcal{T}_F^5$, where  $c_q = 4\mathcal{A}4! \approx 9.8 \times 10^{-10} \, \text{GeV}^{-4}$ . Utilizing equation \eqref{eq:H(T)} at  $\mathcal{T} = \mathcal{T}_F$ , we find
\begin{equation}
     \mathcal{T}_F = \left( \frac{\pi^2 g_*}{90 M^2_{p} c^2_q} \right)^{1/6} \approx 0.6 \, \text{MeV}
\end{equation}

Given $H(\mathcal{T}_F) = c_q \mathcal{T}^5_F$ , we derive  $\Delta H = 5c_q T^4_f \Delta \mathcal{T}_F$. Substituting $\Delta H$  from equation \eqref{eq:deltaH}, we obtain

\begin{equation}\label{eq:deviation}
     \frac{\Delta \mathcal{T}_F}{\mathcal{T}_F} \approx \frac{\rho_{DE}}{\rho_r} \frac{H_{GR}}{10c_q \mathcal{T}^5_F}. 
\end{equation}

Taking into account the observational constraint, we arrive at the expression constraining the freeze-out temperature as

\begin{equation}\label{eq:bbnbound}
    \left| \frac{\Delta \mathcal{T}_F}{\mathcal{T}_F} \right| < 4.7 \times 10^{-4}.
\end{equation}

\section{Constraining $f(T)$ models} \label{sec4}
 From theoretical and observational aspects of BBN, one can derive constraints on a given cosmological model. A cosmological model is presumed to be viable only if it meets the conditions set by BBN. Therefore, in this section, we shall compare various $f(T)$ gravity models with BBN calculations.
 
\subsection{Hybrid exponential model}
We begin considering the Hybrid exponential $f(T)$ model having the form
\begin{equation}\label{eq:model1}
    f(T)=T \left(e^{\frac{n  T_0}{T}}-1\right),
\end{equation}
with $n$ being the free parameter. The model, in similar lines with \citep{Anagnostopoulos:2021ydo}, has proved its worth by providing a better fit to the observational datasets than the $\Lambda CDM$ model. The model reduces to TEGR with the limit $n=0$. The dark energy density \eqref{eq:rhode} for this model can be calculated as
\begin{equation}\label{eq:rhodem1}
\rho_{DE}=\frac{1}{2 \,{M_p}^2} \left( e^{\frac{n  H_0^2}{H^2}} \left(12 \, n \, H_0^2-6 H^2\right)+6 H^2 \right).
\end{equation}
Using the above expression in the first motion equation \eqref{eq:motion1}, we obtain 
\begin{equation}
e^{\frac{n  H_0^2}{H^2}} \left(H^2-2 n  H_0^2\right)=H_0^2 \Omega_{m0}  (z+1)^3
\end{equation}
where $\Omega_{m0}= \rho_{m0}/3{H_0}^2$, is the present density parameter. 

Again we use \eqref{eq:rhodem1} to find the deviation in freeze out temperature from \eqref{eq:deviation},

\begin{equation}\label{eq:Tfm1}
\frac{\Delta \mathcal{T}_F}{\mathcal{T}_F}=\frac{e^A \left(180 \,n \, H_0^2 \,{M_p}^2-\pi ^2 \,g \,{\mathcal{T}_F}^4\right)+\pi ^2 \,g\, {\mathcal{T}_F}^4}{30 \sqrt{10\,g} \,\pi \, c_q\, {M_p}^5 \,{\mathcal{T}_F}^7 },
\end{equation}
where $A={\frac{90 n  H_0^2 {M_p}^2}{\pi ^2 g {\mathcal{T}_F}^4}}$. To constrain the model using the bound on deviation, the values for the fixed parameters are incorporated as $M_p=1.22 \times 10^{19}\,\,GeV$ , $\mathcal{T}_F=0.0006\,\,GeV$, $ c_q=9.8\times 10^{-10} GeV^{-4} $ and $g \sim 10$. Moreover, the present value of the Hubble parameter is considered as $H_0 = 1.47\times 10^{-42\;}GeV$ which is equivalent to the value from Planck 2018 results i.e., $H_0=67.2$ km $s^{-1}$ Mp$c^{-1}$.

\begin{figure}
    \centering
    \includegraphics[width=\linewidth]{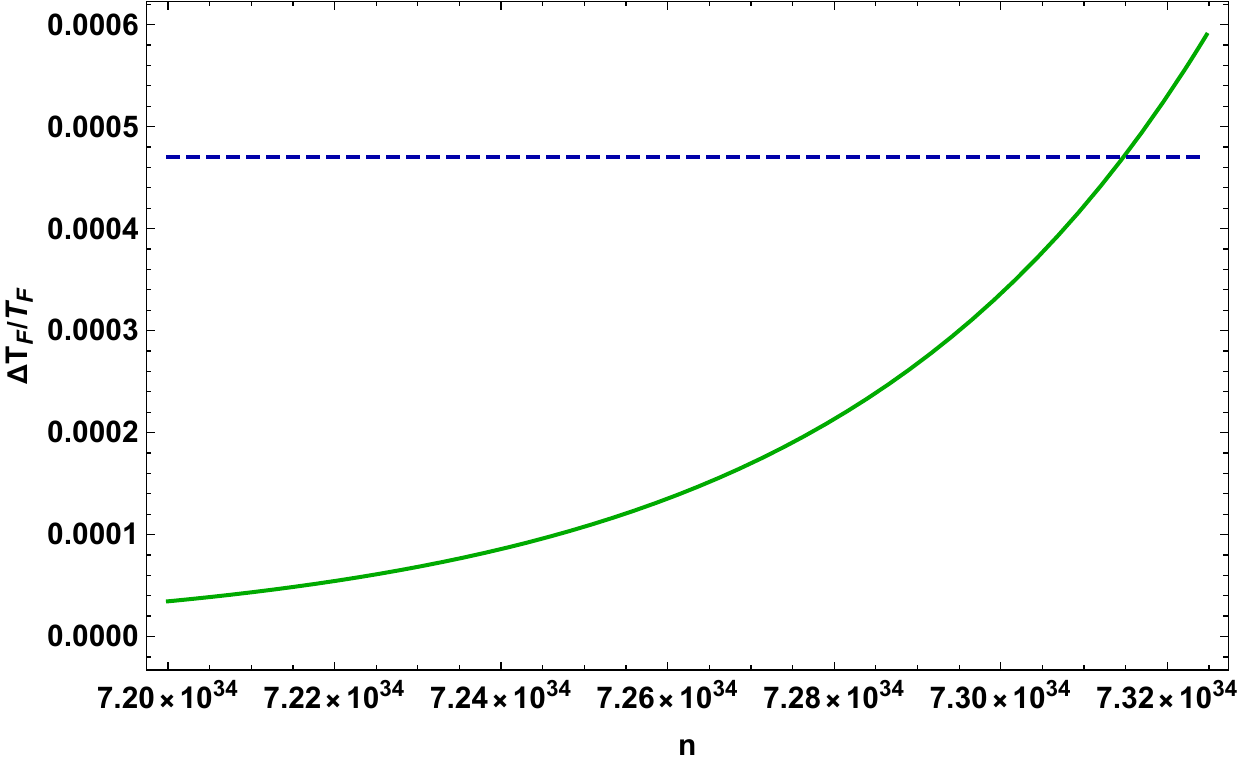}
    \caption{$n$ vs $\Delta \mathcal{T}_F/\mathcal{T}_F$ for the Hybrid exponential model. The blue dashed line represents the upper bound of $\Delta \mathcal{T}_F/\mathcal{T}_F$. }
    \label{fig:bbn1}
\end{figure}

We depict from \autoref{fig:bbn1} that the model satisfies the BBN bound \eqref{eq:bbnbound} for $n \leq 7.3147 \times 10^{34}$. The blue dashed curve represents the upper bound of the deviation $\left(\frac{\Delta \mathcal{T}_F}{\mathcal{T}_F}=0.00047\right)$. As the curve is asymptotic to zero towards the left, the domain we obtain for the model parameter is unbounded below. 

\subsection{Hybrid tangent hyperbolic model}
The Hybrid tangent hyperbolic model is motivated from \citep{Wu:2010av} because it allows phantom divide crossing line for the effective equation of state (EoS) parameter and corroborates with the recent observational datasets. The model consists of a tangent hyperbolic and a power law term that has  the following functional form 

\begin{equation}\label{eq:model2}
    f(T)=\lambda T_0 \,\left(\frac{T}{T_0}\right)^a \tanh\left(\frac{T_0}{T}\right),
\end{equation}
with $a$ and $\lambda$ being the free parameters. The model reduces to TEGR with the limit $\lambda=0$. The dark energy density \eqref{eq:rhode} for this model reads
\begin{multline}\label{eq:rhodem2}
    \rho_{DE}=\frac{\lambda \, M_p^2}{2} \, E ^{2n-2}\left(12\,H_0^2 \, \text{sech}\left(\frac{1}{E}\right)^4+
    \right. \\\left.
    6H^2\,(1-2n)\, \tanh\left(\frac{1}{E}\right)^2\right),
\end{multline}
where $E=H/H_0$. Using the above expression in the first motion equation \eqref{eq:motion1}, we obtain 
\begin{multline}
  H^2=\lambda \left(\frac{H}{H_0}\right)^{2n-2} \left(2 {H_0}^2 \text{sech}\left(\frac{H_0}{H}\right)^{4}+ \right.\\\left.
  H^2(1-2n) \tanh \left(\frac{H_0}{H}\right)^{2}   \right)+\Omega_{m0}(1+z)^3.
\end{multline}

Considering the present time scenario in the above equation, one can obtain the dependency between the model parameters as
\begin{equation}\label{eq:lambda}
    \lambda=\frac{1- \Omega_{m0}}{2 \, \text{sech}(1)+(1-2n)\, \tanh(1)}.
\end{equation}
Using \eqref{eq:rhodem2} in the theoretical expression for the deviation in freeze-out temperature \eqref{eq:deviation}, we get

\begin{equation}\label{eq:Tfm2}
\begin{split}
    \frac{\Delta \mathcal{T}_F}{\mathcal{T}_F}= &\frac{1}{c_q \,g\, {\mathcal{T}_F}^9 \sqrt{\frac{90}{C}} (-2 n \,\tanh (1)+\tanh (1)+2 \, \text{sech}(1))}\times \\& \left[{H_0} \,3^{1-2 n} \,10^{-n-\frac{1}{2}} \,\pi ^{2 n-4}\, (1-\Omega_{m0})\left(\frac{90}{\pi^2 C}\right)^n \times \right.\\&\left.\left(\pi ^2 g (1-2 n)\, {\mathcal{T}_F}^4\, \tanh \left(C\right)+180 {H_0}^2 {M_p}^2 \text{sech}^2\left(C\right)\right)\right],
\end{split}
\end{equation}
where $C=\frac{90 \, {H_0}^2 \, {M_p}^2}{\pi ^2 \, g \, {\mathcal{T}_F}^4}$.

\begin{figure}
    \centering
    \includegraphics[width=\linewidth]{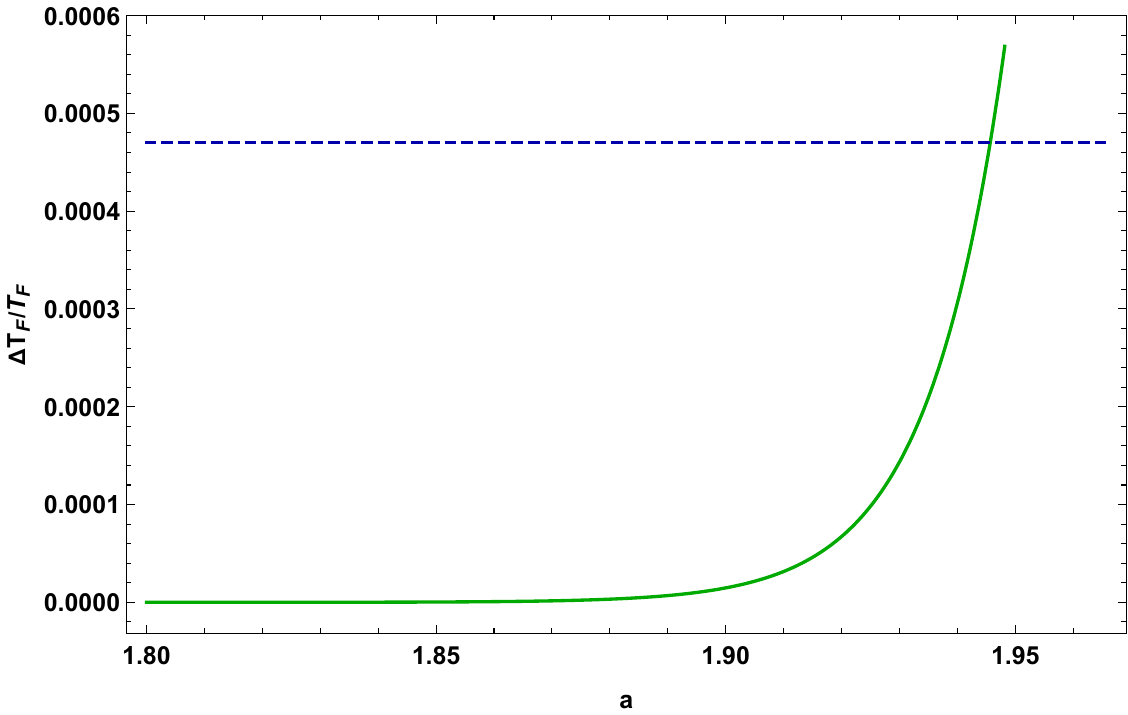}
    \caption{$a$ vs $\Delta \mathcal{T}_F/\mathcal{T}_F$ for the Hybrid tangent hyperbolic model. The blue dashed line represents the upper bound of $\Delta \mathcal{T}_F/\mathcal{T}_F$.}
    \label{fig:bbn2}
\end{figure}

We depict from \autoref{fig:bbn2} that the model satisfies the BBN bound \eqref{eq:bbnbound} for $a \leq 1.94566$. The values of fixed parameters are considered the same as the first model. By straightforward calculation, one can obtain the range for the other model parameter $\lambda$ from \eqref{eq:lambda}. The blue dashed curve represents the upper bound of the deviation $\left(\frac{\Delta \mathcal{T}_F}{\mathcal{T}_F}=0.00047\right)$. Similar to the first model, the curve is asymptotic to zero towards the left, hence the domain for the model parameter is unbounded below.

\section{Verification of the models in late-time using observational datasets}\label{sec5}
\subsection{Datasets}
To develop a cosmological model that viably explains the evolution of the Universe, we need to find valid ranges of space for free parameters. This is achieved through the Bayesian approach along with the Markov Chain Monte Carlo (MCMC) methodology. For this purpose, in our present work, we use Cosmic Chronometers and Gamma-Ray-Bursts datasets.

\subsubsection{Cosmic Chronometers (CC)} 
Using the differential aging technique, the CC method deduces the Hubble rate by studying ancient, quiescent galaxies that are closely positioned in redshift. This approach is based on defining the Hubble rate $ H = -\frac{1}{1+z}\frac{dz}{dt}$  for a FLRW metric. The CC method is significant because it can determine the Hubble parameter independently of any specific cosmological model. Based on various surveys \citep{Jimenez:2003iv,Simon:2004tf,Stern:2009ep,Moresco:2012jh,Zhang:2012mp,Moresco:2015cya,Moresco:2016mzx,Ratsimbazafy:2017vga}, we utilized 34 CC data points in this study within the redshift range of  $0.1 <z< 2$. 

\subsubsection{Gamma-Ray-Bursts (GRBs)}
GRBs are the most intense explosions in the Universe, having enormous energy. It is possible to observe these cosmic phenomena at comparatively higher redshifts. The correlation between $\nu F_{\nu}$, $E_{p,i}$, and $E_{\text{iso}}$ is given by
\begin{equation}
    \log_{10} \left( \frac{E_{\text{iso}}}{1 \, \text{erg}} \right) = b + a \log_{10} \left( \frac{E_{p,i}}{300 \, \text{keV}} \right),
\end{equation}
where $\nu F_{\nu}$ represents the rest-frame spectrum peak energy, $E_{\text{iso}}$ is the isotropic-equivalent radiated energy \citep{Amati:2002ny}, $a$ and $b$ are constants, and $E_{p,i}$ denotes the spectral peak energy in the cosmological rest frame of the GRBs. $E_{p,i}$ correlates with the observer-frame quantity $E_p$ through its relation being derived from $E_p$ as $E_{p,i} = E_p(1 + z)$. $E_{iso}$ can be calibrated using the luminosity distance $d_L$ and bolometric fluence $S_{\text{bolo}}$ which is related as
\begin{equation}
    E_{\text{iso}} = 4\pi {d_L}^2(z, \text{cp}) S_{\text{bolo}} (1 + z)^{-1}.
\end{equation}
In the present work, we analyze $162$ data points for the redshift ranging from $0$ to $9.3$ \citep{Demianski:2016zxi}. This offers the possibility of studying beyond the Supernovae observation and covering a higher redshift.

To find the best-fit parameter space, we maximize the likelihood $\mathcal{L}\propto e^{ -\frac{\chi^2}{2} }$, where $\chi^2$ function is defined by,
\begin{equation}
    \chi^2=\Delta V^T (C^{-1} )\Delta V.
\end{equation}
Here, $C$ denotes the covariance matrix of uncertainties, and $\Delta V$ is the difference between the theoretical and observational values of the key physical quantity. For instance, $V$ signifies the Hubble parameter for CC and the distance modulus function to that of GRBs. 

\subsection{Results and Interpretation}
In the previous section, we restricted the model parameters of two highly motivated models through BBN constraints. This section is dedicated to test the credibility of the said models in various epochs while staying within the limits.

\begin{figure}
    \centering
    \includegraphics[width=\linewidth]{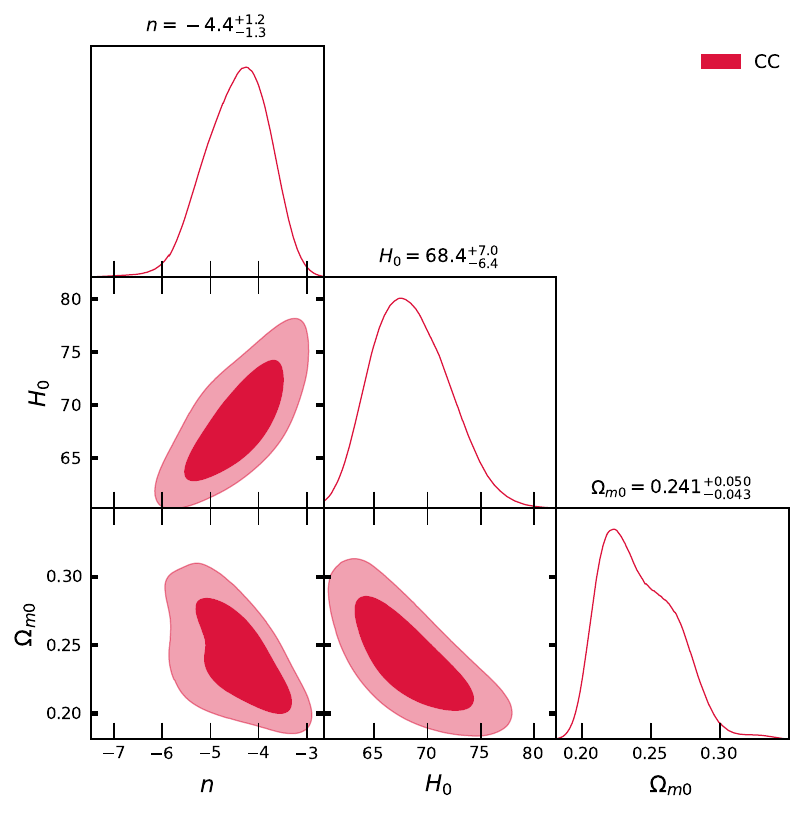}
    \caption{Likelihood probability distributions and 2D contours for the Hybrid exponential model obtained from MCMC analysis of the CC dataset. The dark-shaded region represents $1 \sigma$ CL and the light-shaded region represents $2 \sigma$ CL. }
    \label{fig:c1}
\end{figure}

\begin{figure}
    \centering
    \includegraphics[width=\linewidth]{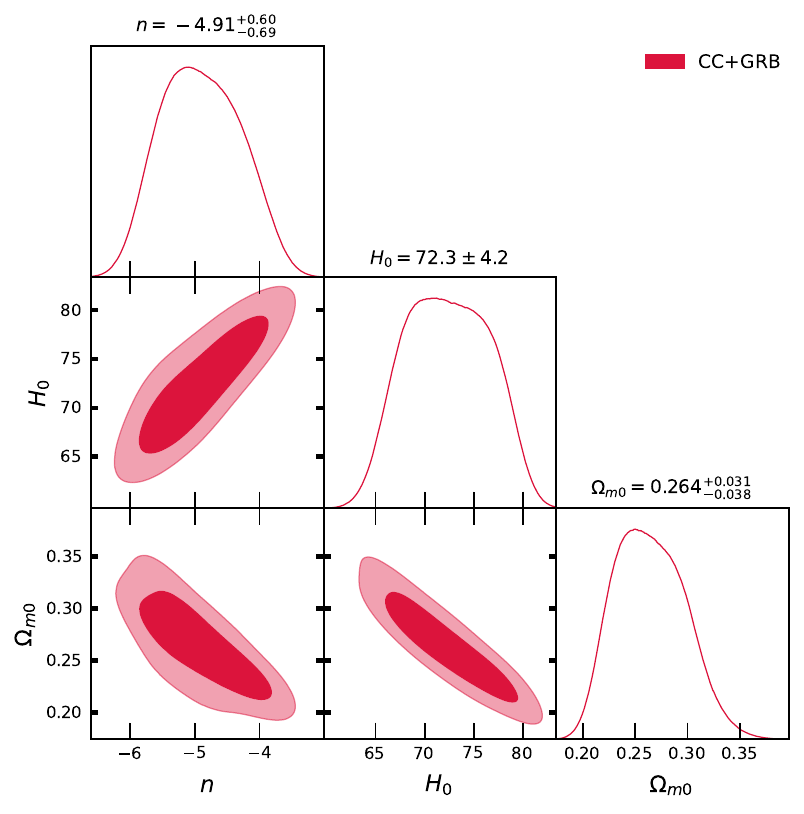}

        \caption{Likelihood probability distributions and 2D contours for the Hybrid exponential model obtained from MCMC analysis of the CC + GRB dataset. The dark-shaded region represents $1 \sigma$ CL and the light-shaded region represents $2 \sigma$ CL.}
    \label{fig:c2}
\end{figure}

\begin{figure}
    \centering
    \includegraphics[width=\linewidth]{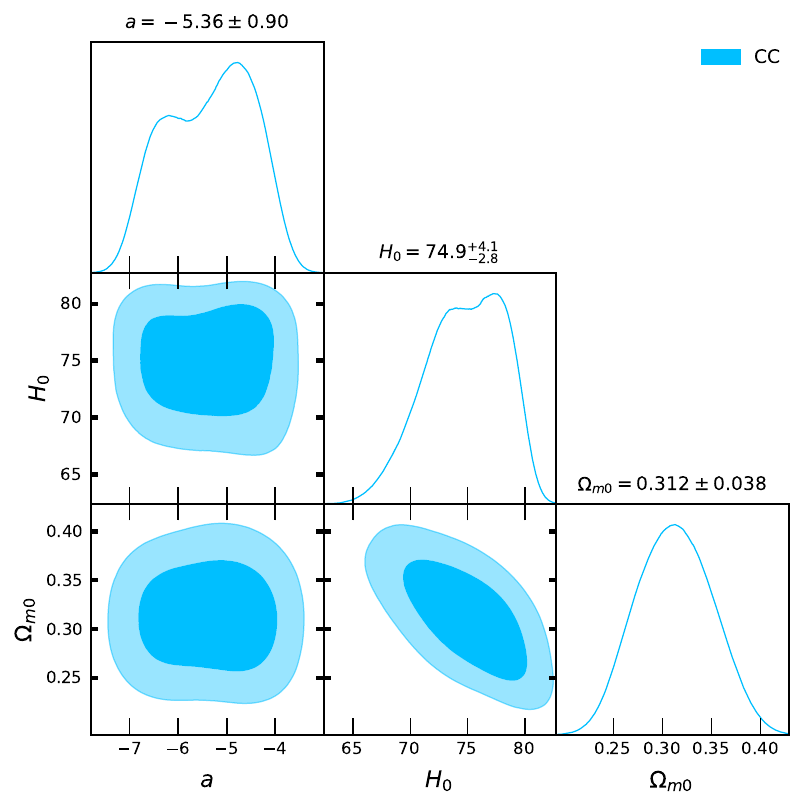}
    \caption{Likelihood probability distributions and 2D contours for the Hybrid tangent hyperbolic model obtained from MCMC analysis of the CC dataset. The dark-shaded region represents $1 \sigma$ CL and the light-shaded region represents $2 \sigma$ CL.}
    \label{fig:c3}
\end{figure}

\begin{figure}
    \centering
    \includegraphics[width=\linewidth]{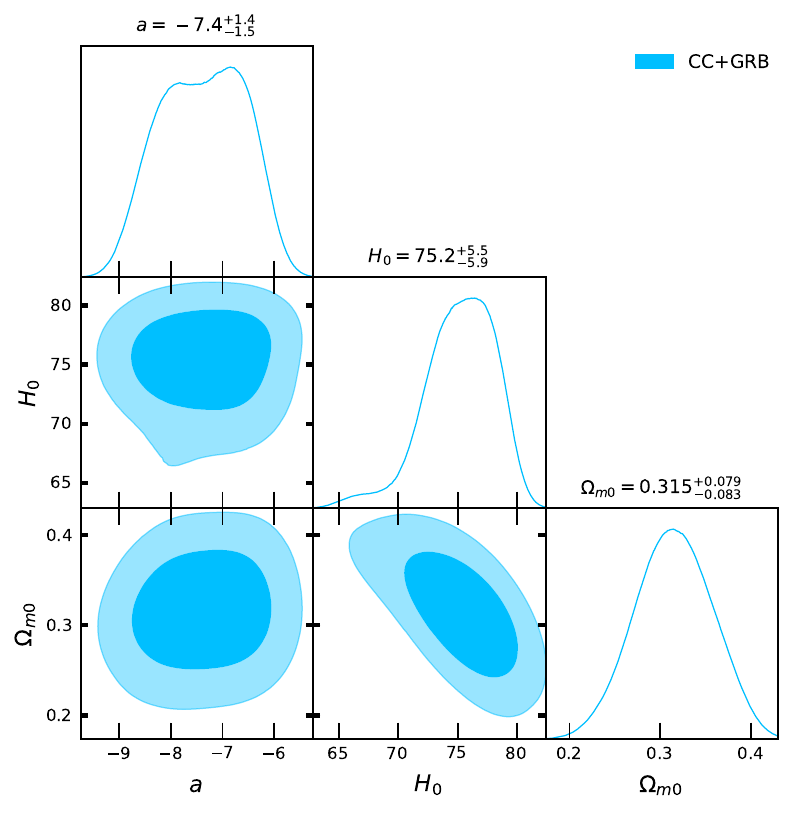}
    \caption{Likelihood probability distributions and 2D contours for the Hybrid tangent hyperbolic model obtained from MCMC analysis of the CC + GRB dataset. The dark-shaded region represents $1 \sigma$ CL and the light-shaded region represents $2 \sigma$ CL.}
    \label{fig:c4}
\end{figure}

\begin{figure}
    \centering
    \includegraphics[width=\linewidth]{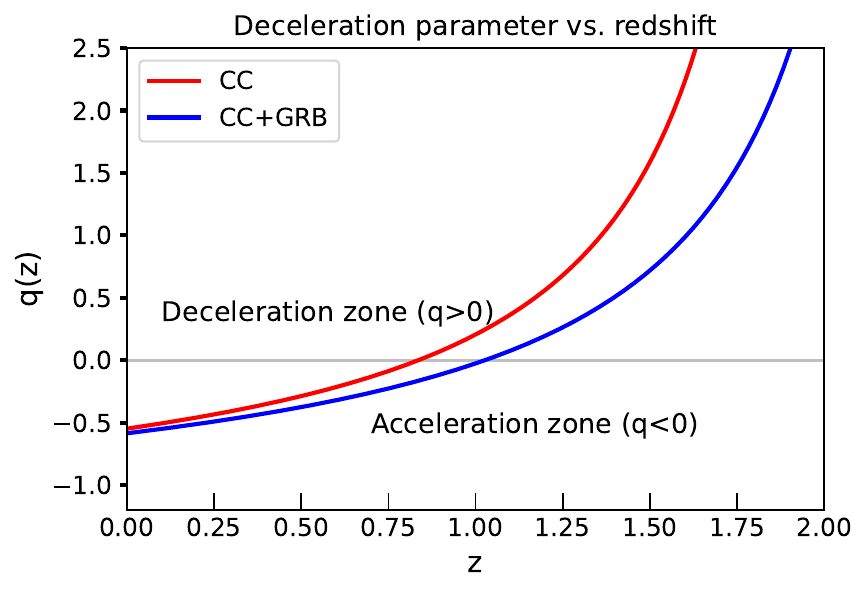}
    \caption{Deceleration parameter (constrained from CC and CC+GRB) vs redshift for the Hybrid exponential model, indicating the phase transition.}
    \label{fig:DP1}
\end{figure}

\begin{figure}
    \centering
    \includegraphics[width=\linewidth]{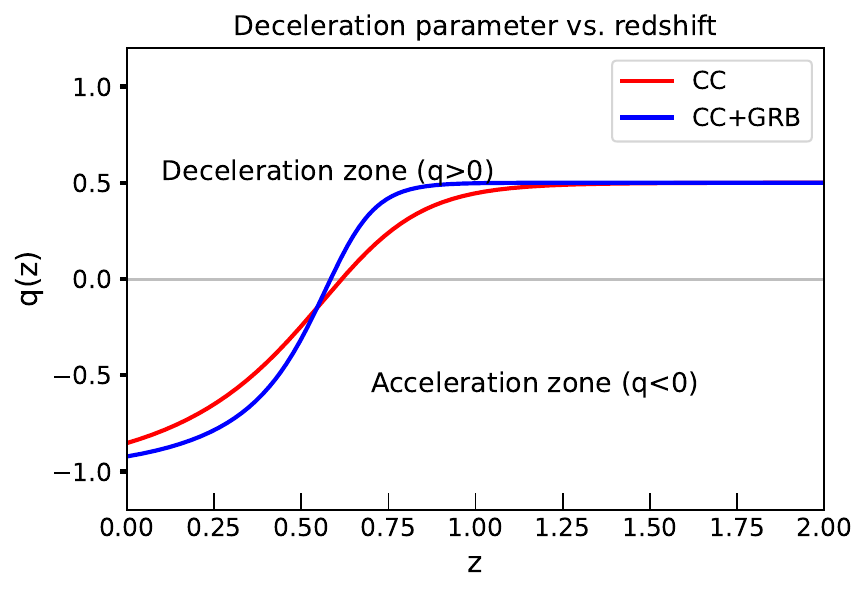}
    \caption{Deceleration parameter (constrained from CC and CC+GRB) vs redshift for the Hybrid tangent hyperbolic model, indicating the phase transition.}
    \label{fig:DP2}
\end{figure}

We start by performing an MCMC analysis to constrain the free parameters using the CC and GRB datasets. It can be observed in \autoref{fig:c1}-\ref{fig:c4} that the model parameter ranges obtained from the sampling lie well within the BBN range which confirms that the models are suitable candidates to explain the early time as well as late-time behavior of the Universe. However, agreement with all the intermediating epochs is another essential requirement for any theory to be considered as an alternative to GR. In this context, we study the behavior of the deceleration parameter for both our models. For the hybrid exponential model, one can observe in \autoref{fig:DP1} that the parameter transits from the deceleration to the acceleration zone at redshifts mentioned in \autoref{table1}. On the other hand, for the hybrid tangent hyperbolic model in \autoref{fig:DP2}, the $\it{q}$-profile shows a better trend in the phase transition with more accurate redshifts. The present values of the deceleration parameter and the transition redshifts summarized in \autoref{table1} are found to be suitably aligned with the recent observational data. 

\begin{figure}
    \centering
    \includegraphics[width=\linewidth]{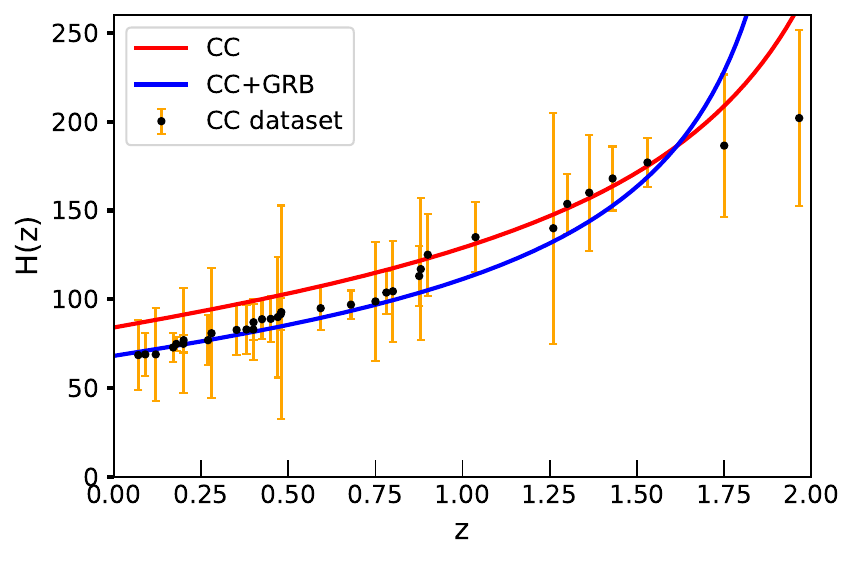}
    \caption{The Hubble function (constrained from CC and CC+GRB) 
    against redshift and error bars of CC dataset for the Hybrid exponential model.}
    \label{fig:HP1}
\end{figure}

\begin{figure}
    \centering
    \includegraphics[width=\linewidth]{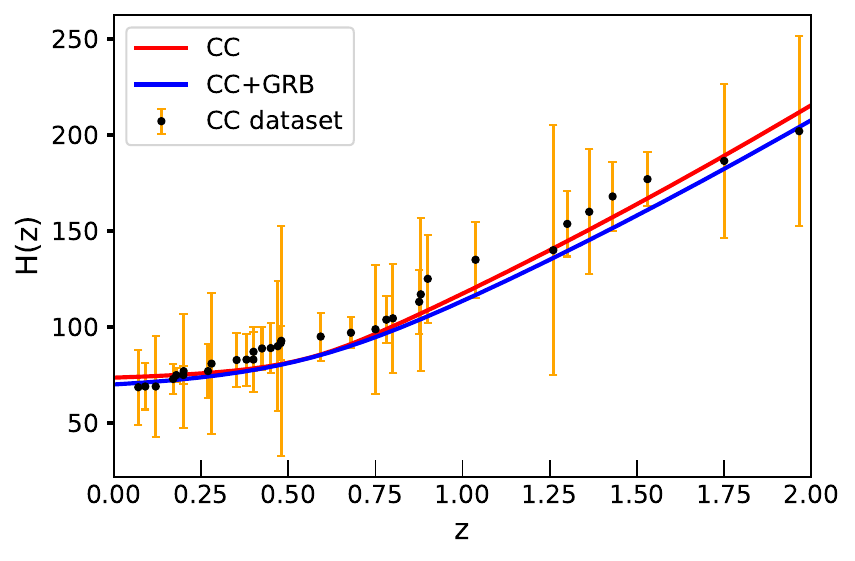}
    \caption{The Hubble function (constrained from CC and CC+GRB) 
    against redshift and error bars of CC dataset for the Hybrid tangent hyperbolic model.}
    \label{fig:HP2}
\end{figure}

\begin{figure}
    \centering
    \includegraphics[width=\linewidth]{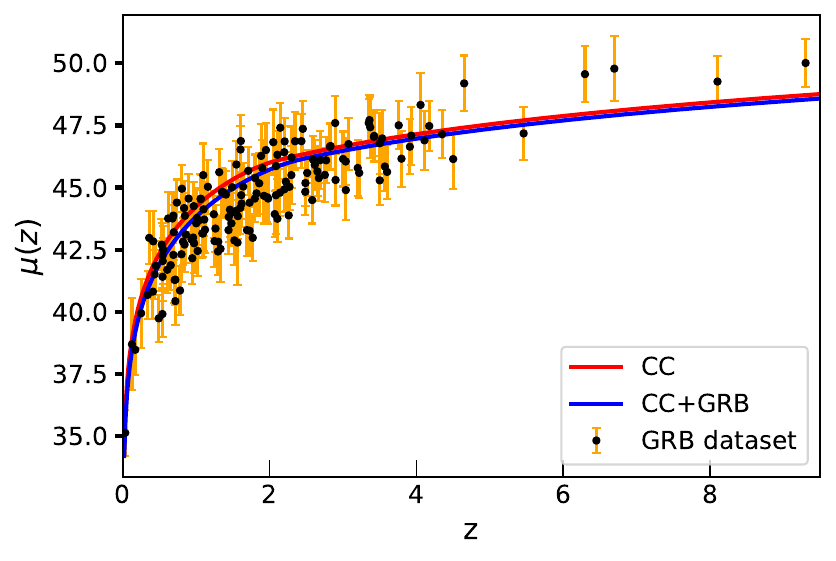}
    \caption{The distance modulus function (constrained from CC and CC+GRB) 
    against redshift and error bars of GRB dataset for the Hybrid exponential model.}
    \label{fig:mu1}
\end{figure}

\begin{figure}
    \centering
    \includegraphics[width=\linewidth]{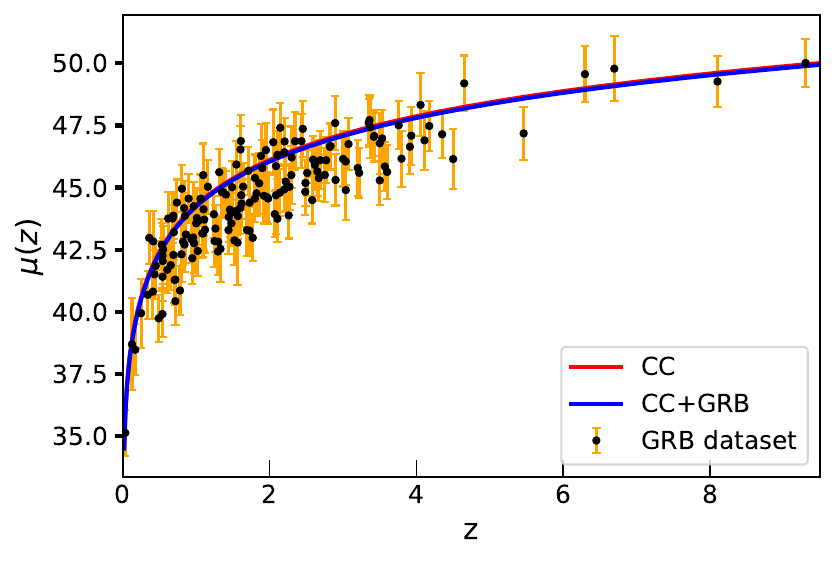}
    \caption{The distance modulus function (constrained from CC and CC+GRB) 
    against redshift and error bars of GRB dataset for the Hybrid tangent hyperbolic model.}
    \label{fig:mu2}
\end{figure}

Furthermore, the Hubble and distance modulus functions are confronted with the 34 data points of CC and 162 data points of GRB datasets, respectively. The GRB data is particularly considered because of its capability to explore the higher redshifts. For both models, we find the functions are in nice agreement with the data points (See \autoref{fig:HP1}-\ref{fig:mu2}). The common ranges obtained for the model parameters from considering early and late-time analysis are summarized in \autoref{table2}.



\begin{table}
 \centering
 \caption{Summary of the transition redshift $(z_t)$ and present value of deceleration parameter $(q_0)$ obtained from $q(z)$ of both models.
 }
 
 \label{table1}
 \resizebox{\linewidth}{!}{
    \begin{tabular}{|cc|c|c|c}
    
 \cline{1-4}
\multicolumn{1}{|c|} {\it{Model}} & \it{Dataset} & {$z_t$} & $q_0$ \\ \cline{1-4}
\multicolumn{1}{ |c  }{\multirow{2}{*}{Hybrid Exponential} } &
\multicolumn{1}{ |c| }{CC} & 0.838 & -0.548 &      \\ 
\multicolumn{1}{ |c  }{}                        &
\multicolumn{1}{ |c| }{CC + GRB} & 1.026 & -0.585 &     \\ \cline{1-4}
\multicolumn{1}{ |c  }{\multirow{2}{*}{Hybrid Tangent Hyperbolic} } &
\multicolumn{1}{ |c| }{CC} & 0.617 & -0.852 &  \\ 
\multicolumn{1}{ |c  }{}                        &
\multicolumn{1}{ |c| }{CC + GRB} & 0.585 & -0.921 & \\ \cline{1-4}   
    \end{tabular}}
\end{table}

\begin{table}
 \centering
 \caption{Summary of the ranges obtained from both early and late time for the $f(T)$ models. In the first row, the results of model parameter $n$ of the Hybrid Exponential model are presented while in the second row, the results of model parameter $a$ of the Hybrid Tangent Hyperbolic model are presented.}
 
 \label{table2}
 \resizebox{\linewidth}{!}{
    \begin{tabular}{|cc|c|c|c}
    
 \cline{1-4}
\multicolumn{1}{|c|} {\it{Model Parameter}} & \it{Observation} & \it{Range} & \it{Common Range}  \\ \cline{1-4}
\multicolumn{1}{ |c  }{\multirow{2}{*}{$n$}} &
\multicolumn{1}{ |c| }{CC} & $(-5.7,-3.2)$ &      \\ 
\multicolumn{1}{ |c  }{}  
             &
\multicolumn{1}{ |c| }{CC + GRB} & $(-5.6,-4.31)$ &  $(-5.6,-4.31)$ &   \\ 
\multicolumn{1}{ |c  }{} 
             &
\multicolumn{1}{ |c| }{BBN} & $(-\infty,7.3147 \times 10 ^{34})$ &    \\ \cline{1-4}
\multicolumn{1}{ |c  }{\multirow{2}{*}{$a$} } &
\multicolumn{1}{ |c| }{CC} & $(-6.26,-4.46)$ &   \\ 
\multicolumn{1}{ |c  }{}                        &
\multicolumn{1}{ |c| }{CC + GRB} & $(-8.9,-6)$ & $(-6.26,-6)$ & \\ 
\multicolumn{1}{ |c  }{}
              &
\multicolumn{1}{ |c| }{BBN} & $(-\infty,1.94566)$ & \\ \cline{1-4}   

    \end{tabular}}
\end{table}

\section{Concluding Remarks}\label{sec6}

To execute this work, we adopted a new approach that includes different evolutionary phases of the Universe starting from the nucleosynthesis era. The challenging outcomes of this work raise curiosity to know more about the assumed teleparallel models. The presence of $\it{T}$ term in the denominator makes both the functions approach zero in early times $(T \rightarrow \infty)$. This indicates that without any limits on the model parameters, the models reduce to TEGR at the time of the Big Bang, which solves the cosmological constant problem. A similar line of reasoning can be followed for the late-time as well, the coupling with polynomial $\it{T}$ term makes it GR-like when $T \rightarrow 0$. 

To find a proper justification for whether the models are eligible candidates to connect the early-time and late-time epochs, we commenced from the primordial era by constraining them using BBN observations. The model parameters are restricted as $n \leq 7.3147 \times 10^{34}$ for the exponential model and $a \leq 1.94566$ for the tangent hyperbolic model. Two widely used datasets CC and GRB have been used to constrain the models further by performing the MCMC technique. It is to be noted that the ranges of the model parameters align with the range obtained from BBN and the other cosmological parameters are in agreement with the observational data \citep{Planck:2018vyg,Mandal:2023bzo}. 

\textit{Deceleration parameter}, the most important tool to show phase changes in the intermediate epochs, is then studied to observe the behavior of our models in the transitional times. We find that both models show the transition from the deceleration zone to the acceleration zone. The redshift at which the alteration is occurring is slightly higher \citep{Rani:2015lia,Naik:2023yhl} for the Hybrid exponential model constrained by CC+GRB, while it is around the most widely accepted value for the remaining three \citep{Cunha:2008ja}. For the hybrid tangent hyperbolic model, the deceleration parameter approaches $0.5$ for higher redshift which coincides with the behavior of the $\Lambda$CDM model. However, for the hybrid exponential model, a definitive intermediate epoch cannot be observed as it deviates after a certain redshift. Nevertheless, in both models, the present values of the deceleration parameter confirm the ongoing accelerated expansion of the Universe. In addition, the constrained teleparallel $f(T)$ models are found to be viable in the early phase of the Universe and they align with the data points of the CC and GRB dataset, which is verified using $H(z)$ and $\mu(z)$ functions. 

Thus by joining the pieces, we conclude that both the hybrid exponential and hybrid tangent hyperbolic models can fairly explain the physical aspects of the dynamics of the Universe. In particular, the hybrid tangent hyperbolic model can be considered a fine alternative to GR. 

\section*{Data availability} No new data were generated or analysed in support of this research.

\section*{Acknowledgement}
  NSK, and VV acknowledge DST, New Delhi, India, for its financial support for research facilities under DST-FIST-2019. 
 SSM acknowledges the Council of Scientific and Industrial Research (CSIR), Govt. of India for awarding Junior Research fellowship (E-Certificate No.: JUN21C05815). PKS acknowledges Science and Engineering Research Board, Department of Science and Technology, Government of India for financial support to carry out Research project No.: CRG/2022/001847.  We are very much grateful to the honorable referee and to the editor for the illuminating suggestions that have significantly improved our work in terms
of research quality, and presentation.

\label{lastpage}
\bibliographystyle{mnras}
\nocite{}
\bibliography{main}
\end{document}